\documentclass[a4paper,12pt]{article}
\usepackage{amsfonts}

\input{tcilatex}

\begin{document}

\title{Additivity of Entangled Channel Capacity for Quantum Input States}
\author{V.P.Belavkin and X.Dai \\
School of Mathematical Sciences\\
University of Nottingham, UK}
\maketitle

\abstract{An elementary introduction into algebraic approach 
to unified quantum information theory and operational approach to
quantum entanglement as generalized encoding is given.
After introducing compound quantum state and two types
of informational divergences, namely, Araki-Umegaki 
(a-type) and of Belavkin-Staszewski (b-type)
quantum relative entropic information,
this paper treats two types of quantum mutual information via entanglement
and defines two types of corresponding 
quantum channel capacities as the supremum via
the generalized encodings. It proves the additivity property of
quantum channel capacities via entanglement, which extends
the earlier results of V. P. Belavkin to products 
of arbitrary quantum channels for quantum relative
entropy of any type.}

\section{Introduction}

Unlike classical channels, quantum channels can have several different
capacities (e.g. for sending classical information or quantum information,
one-way or two-way communication, prior or via entanglement, etc.). Well,
the problem of characterizing in general the capacity of a noisy quantum
channel is unsolved although several attempts have been made to define a
quantum analog of Shannon mutual information (see the conceptions of
coherent information [30] or von Neumann mutual entropy [20, 21, 24]).
Unfortunately most of these attempts do not give a satisfactory solution
because the defined quantities fail to preserve such important property of
classical informational capacity as additivity and some do not have even
monotonicity property. This paper is based on the operational entanglement
approach to quantum channel capacity suggested in [10-12], which is free of
the above difficulties due to the enlargement of the class of input
encodings, including the encodings via entanglement for one-way
communication.

Quantum entanglement is a uniquely quantum mechanical resource that plays a
key role, along with the celebrating paper [28] of Einstein-Podolsky-Rosen,
in many of the most interesting applications of quantum information and
quantum computation, such as, Quantum entanglement is extensively used in
teleporting an unknown quantum state via dual classic and
Einstein-Podolsky-Rosen channels in the subject of quantum teleportation
[1], quantum cryptography was investigated based on Bell's theorem [2],
quantum noiseless coding theorem appeared [3] as a quantum analogous of
Shannon's classical noiseless coding theorem. This paper will concentrate on
application of quantum entanglement to quantum source entropy and quantum
channel capacity in the subject of quantum information.

Recently tremendous effort has been made to better understand the properties
of quantum entanglement as a fundamental resource of nature. Although there
is as yet no complete understanding and proof of physical realizability of
quantum entanglement for quantum technologies, a theoretical progress has
been made in understanding this strange property of quantum mechanics, for
example, mathematical aspects of quantum entanglement are extensively
studied V. P. Belavkin [27] described the dynamical procedure of quantum
entanglement in terms of transpose-completely positive maps in the subject
of quantum decoherence and stochastic filtering theory, V. P. Belavkin and
M. Ohya [11,12] initiated mathematical study of quantum entanglement as
truly quantum couplings from an operational view in algebraic approach,
Peter Levay [15] investigated geometry of quantum entanglement for two
qubits (quantum entanglement of two qubits corresponds to the twisting of
the bundle), R. Penrose [16] treated quantum entanglement via spinor
representation in the subject of mathematical physics, Peter Levay [17]
investigated twistor geometry of quantum entanglement for three qubits still
in mathematical physics. This paper will follow [10,11,12] to treat with
quantum entanglement in algebraic approach.

Using the operational treatment of entanglement as "true quantum" encoding
V. P. Belavkin and M. Ohya [10,11,12] introduced quantum conditional entropy
of the entangled compound state related to the product of marginal states
which is positive and obeys all natural properties of the classical
conditional entropy as the relative conditional/unconditional entropy of a
compound state. They studied its relation to the mutual information as the
informational divergence (relative informational entropy) of the compound
state with respect to the product of its marginal states in the sense of
Lindblad, Araki and Umegaki [14,4,5]. This quantum mutual information leads
to an entropy bound of quantum mutual information and quantum channel
capacity via entanglement (entanglement-assisted quantum capacity introduced
in [20,21]), which considered the mutual information of input-output state
of quantum channel. Also V. P. Belavkin and P. Staszewski [25] investigated
C*-algebraic generalization of relative and conditional entropy including
two types of quantum relative entropy, such as Araki-Umegaki type and
Belavkin-Staszewski type, and even more general informational divergencies
which meet natural axiomatic properties of relative information were studied
in quantum information in [29].

Based on the combination of these two original ideas, after introducing
compound quantum state and two types of quantum relative entropy, namely
Araki-Umegaki type and Belavkin-Staszewski type, this paper treats two types
of quantum mutual information via entanglement in algebraic approach and
corresponding quantum channel capacities via entanglement in operational
approach. It proves additivity property of quantum channel capacities via
entanglement, which extends the results of V. P. Belavkin [10,10a] to
products of arbitrary quantum channel and to quantum relative entropy of not
only Araki-Umegaki type but also Belavkin-Staszewski type.

The rest of this paper is organized as follows: section two and three
introduce related notion of quantum mechanics, such as quantum state and
quantum entanglement respectively; section four introduces two types of
quantum relative entropy via entanglement; section five introduces quantum
channel capacity via entanglement and show additivity of quantum channel
capacity via entanglement; final section contributes to conclusion and
further problems.

\section{Quantum States in Algebraic Approach}

This subsection is a brief mathematical review of Quantum State in Quantum
Mechanics in a discrete algebraic approach. Anyone can turn to [6] for
general physical review, or [7] for mathematical foundations of Quantum
Mechanics, [8,9] for a brief review of Quantum Mechanics Principles in
quantum information and computation.

In order to keep a closer link with classical information theory, we will
allow for a possibility of having classical-quantum combined systems
described in what follows by discrete non-commutative W*-algebras $\mathcal{A%
}=\left( \mathcal{A}_{i}\right) $ represented by block-diagonal matrices $%
A=[A(i)\delta _{j}^{i}]$ with arbitrary uniformly bounded operators $A\left(
i\right) \in \mathcal{A}_{i}$ on some separable Hilbert spaces $\mathcal{G}%
_{i}$.

Let $\mathcal{H}$ denote the separable Hilbert space of a quantum system,
and $\mathcal{L(H)}$ denote the algebra of all linear bounded operators on $%
\mathcal{H}$, with a decomposable subalgebra $\mathcal{B}\subseteq \mathcal{%
L(H)}$ of elements $B\in \mathcal{B}$ of the block-diagonal form $%
B=[B(j)\delta _{j}^{i}]$, where $B(j)\in \mathcal{L}(\mathcal{H}_{j})$,
corresponding to an orthogonal decomposition $\mathcal{H}=\bigoplus_{j}%
\mathcal{H}_{j}$. Note that any such algebra is weakly closed in $\mathcal{L}%
(\mathcal{H})$, i.e. is a $W^{\ast }$-algebra having a predual space $%
\mathcal{B}_{\ast }$, which can be identified with the trace class subspace
of $\mathcal{B}$ with respect to the pairing%
\[
\left\langle \varsigma |B\right\rangle =\sum_{j}\mathrm{Tr}_{\mathcal{H}%
_{j}}[\varsigma \left( j\right) ^{\dagger }B\left( j\right) ]=\mathrm{Tr}_{%
\mathcal{H}}[B\varsigma ^{\dagger }], 
\]%
where $\varsigma \left( j\right) \in \mathcal{B}_{j}$ are such operators in $%
\mathcal{H}_{j}$ that $\mathrm{Tr}_{\mathcal{H}}\sqrt{\varsigma ^{\dagger
}\varsigma }<\infty $ and $\mathrm{Tr}_{\mathcal{H}}$ is the standard trace
on $\mathcal{B}$ normalized on one dimensional projectors $P_{\psi }=\psi
\psi ^{\dag }$ for $\psi \in \mathcal{H}_{j}$. We now remind the definition
of quantum normal state.

\textbf{Definition 1} A bounded linear functional $\sigma :\mathcal{B}%
\rightarrow \mathbb{C}$ of the form $\sigma (B)=\mathrm{Tr}_{\mathcal{H}%
}[B\varsigma ]$ for a $\varsigma =\varsigma ^{\dagger }\in \mathcal{B}_{\ast
}$ is called the state on $\mathcal{B}$ if it is positive for any positive
operator $B\in \mathcal{B}$ and normalized $\sigma (I)=1$ for the identity
operator $I$ in $\mathcal{B}$. The operator $\varsigma $, uniquely defined
as a positive trace one operator on $\mathcal{H}$, is called density
operator of the state $\sigma $.

Let $\mathcal{G}$ be another separable Hilbert space and $\chi $ be a
Hilbert-Schmidt operator from $\mathcal{G}$ to $\mathcal{H}$ defining a
decomposition $\varsigma =\chi \chi ^{\dag }$ of the state density with the
adjoint operator $\chi ^{\dag }$ from $\mathcal{H}$ to $\mathcal{G}$. We now
equip $\mathcal{G}$ with an isometric involution $J=J^{\dag },J^{2}=I$, the
complex conjugation on $\mathcal{G}$, 
\begin{equation}
J\Sigma _{k}\lambda _{k}\zeta _{k}=\Sigma _{k}\bar{\lambda _{k}}J\zeta
_{k},\forall \lambda _{k}\in \mathbb{C},\zeta _{k}\in \mathcal{G},
\end{equation}%
defining an isometric transposition $\widetilde{A}=JA^{\dag }J=\overline{A}%
^{\dagger }$ on the algebra $\mathcal{L}\left( \mathcal{G}\right) $, where $%
\overline{A}=JAJ$. A normal state $\rho :\mathcal{A}\rightarrow \mathbb{C}$
on the algebra $\mathcal{A}\subseteq \mathcal{L}(\mathcal{G})$ is called
real (or equivalently symmetric) if its density is real, $\overline{\varrho }%
=\varrho $ (or equivalently symmetric, $\widetilde{\varrho }=\varrho $).
Given a state, $J$ can be always chosen in such a way that $\varrho =%
\overline{\varrho }$ as it was done in [10,11], but here we fix $J$ but not $%
\varrho $, and in general we will not assume that $\varrho =\widetilde{%
\varrho }$. Instead, we may assume that the transposition leaves invariant
the decomposable subalgebra $\mathcal{A}\subseteq \mathcal{L}(\mathcal{G})$
such that $\overline{\mathcal{A}}:=J\mathcal{A}J=\mathcal{A}$, however from
the notational and operational point of view, it is preferable to
distinguish the algebra $\mathcal{A}$ from the transposed algebras $%
\widetilde{\mathcal{A}}=\{\widetilde{A}:A\in \mathcal{A}\}=\overline{%
\mathcal{A}}$.

\textbf{Lemma 1} ([10,11,12]) Any normal state $\rho $ on $\mathcal{A}%
\subseteq \mathcal{L}(\mathcal{G})$ can be expressed as 
\begin{equation}
\rho (A)=\mathrm{Tr}_{\mathcal{H}}[\chi \widetilde{A}\chi ^{\dagger }]=%
\mathrm{Tr}_{\mathcal{H}}[A\varrho ],
\end{equation}

\noindent where the density operator $\varrho \in \mathcal{A}_{\ast }$ is
uniquely defined by $\widetilde{\varrho }=\chi ^{\dagger }\chi =\overline{%
\varrho }$ iff $\chi ^{\dagger }\chi \in \widetilde{\mathcal{A}}$.

Thus we have an operational expression $\rho (A)=\left\langle \chi \overline{%
A}\chi ^{\dagger }|I\right\rangle $ of quantum normal state, which is called
standard in the case $\mathcal{G}=\mathcal{H}$ and $\chi =\sqrt{\varsigma }$%
, in which case $\varrho =\overline{\varsigma }$. Generally $\chi $ is named
as the amplitude operator, or simply amplitude given by a vector $\chi =\psi
\in \mathcal{\ H}$ with $\psi ^{\dag }\psi =\Vert \psi \Vert ^{2}=1$ in the
case of one dimensional $\mathcal{G}=\mathbb{C}$, corresponding to the pure
state $\sigma (B)=\psi ^{\dag }B\psi $, where $\chi ^{\dag }$ is the
functional $\psi ^{\dag }$ from $\mathcal{H}$ to complex field $\mathbb{C}$.

\textbf{Remark 1} The amplitude operator $\chi $ is unique up to a unitary
transform in $\mathcal{H}$ as a probability amplitude satisfying the
conditions $\chi ^{\dag }\chi \in \widetilde{\mathcal{A}}$ such that $%
\varrho =\overline{\chi ^{\dag }\chi }$ is positive decomposable trace one
operator $\varrho =\oplus _{i}\varrho (i)$ with the components $\varrho
(i)\in \mathcal{L}(\mathcal{G}_{i})$ normalized as 
\begin{equation}
\mathrm{Tr}_{\mathcal{G}_{i}}\varrho (i)=k(i)\geq 0,\sum_{i}k(i)=1.
\end{equation}%
Therefore we can identify the predual space $\mathcal{A}_{\ast }$ with the
direct sum $\bigoplus \mathcal{T}(\mathcal{G}_{i})\subseteq \mathcal{A}$ of
the Banach spaces $\mathcal{T}(\mathcal{G}_{i})$ of trace class operators in 
$\mathcal{G}_{i}$.

Note that we denote the probability operators $P_{\mathcal{A}}=\varrho \in 
\mathcal{A}_{\ast }$, $P_{\mathcal{B}}=\varsigma \in \mathcal{B}_{\ast }$ as
trace densities of the states $\rho ,\sigma $ defined as the expectations on
the algebras $\mathcal{A},\mathcal{B}$ respectively by the variations of
Greek letters $\rho ,\sigma $ which are also used in [12] for the transposed
(contravariant) density operators $\widetilde{\varrho }\equiv \rho =%
\overline{\varrho },\widetilde{\varsigma }\equiv \sigma =\overline{\varsigma 
}$ with respect to the bilinear pairings $\rho \left( A\right) =\left\langle
A,\rho \right\rangle \equiv \left\langle \overline{\rho }|A\right\rangle ,$ $%
\sigma \left( B\right) =\left\langle B,\sigma \right\rangle \equiv
\left\langle \overline{\sigma }|B\right\rangle $.

We now define an entangled state $\omega $ on the $W^{\ast }$-tensor product
algebra $\mathcal{A}\otimes \mathcal{B}$ of bounded operators on the Hilbert
product space $\mathcal{G}\otimes \mathcal{H}$ by 
\begin{equation}
\mathrm{Tr}_{\mathcal{G}}[\widetilde{A}\chi ^{\dag }B\chi ]=\omega (A\otimes
B)=\mathrm{Tr}_{\mathcal{H}}[\chi \widetilde{A}\chi ^{\dag }B].
\end{equation}

Obviously $\omega $ can be uniquely extended by linearity to a normal state
on the algebra $\mathcal{A}\otimes \mathcal{B}$ generated by all the linear
combinations $C=\Sigma _{k}\lambda _{k}A_{k}\otimes B_{k}$ such that $\omega
(C^{\dag }C)=\mathrm{Tr}_{\mathcal{G}}[X^{\dag }X]\geq 0$, where $X=\Sigma
_{k}\lambda _{k}B_{k}\chi \tilde{A_{k}}$, and $\omega (I\otimes I)=\mathrm{Tr%
}[\chi ^{\dag }\chi ]=1$.

\textbf{Remark 2} The state (4) is pure on $\mathcal{L}(\mathcal{G}\otimes 
\mathcal{H})$, since it is given by an amplitude $\psi \in \mathcal{G}%
\otimes \mathcal{H}$ defined as $(\zeta \otimes \eta )^{\dag }\psi =\eta
^{\dag }\chi J\zeta $, $\forall \zeta \in \mathcal{G},\eta \in \mathcal{H}$,
with the states $\rho $ on $\mathcal{A}$ and $\sigma $ on $\mathcal{B}$ as
the marginals of $\omega $: 
\begin{equation}
\sigma (B)=\omega (I\otimes B)=\mathrm{Tr}_{\mathcal{H}}[B\varsigma ],\;\rho
(A)=\omega (A\otimes I)=\mathrm{Tr}_{\mathcal{G}}[\widetilde{A}\varrho ].
\end{equation}

Therefore, we call the state $\omega $ defined above as a pure entanglement
state for $\mathcal{A}=\mathcal{L}(\mathcal{G})$, $\mathcal{B}=\mathcal{L}(%
\mathcal{H})$.

More general, mixed entangled states for $\mathcal{A}=\mathcal{L}(\mathcal{G}%
)$, $\mathcal{B}=\mathcal{L}(\mathcal{H})$ can be obtained by using a
stochastic amplitude operator $\chi :\mathcal{G}\rightarrow \mathcal{F}%
\otimes \mathcal{H}$.

Given an amplitude operator $\upsilon :\mathcal{F}\longrightarrow \mathcal{G}%
\otimes \mathcal{H}$ on a Hilbert space $\mathcal{F}$ into the tensor
product Hilbert space $\mathcal{G}\otimes \mathcal{H}$ such that $\varpi
:=\upsilon \upsilon ^{\dag }\in \mathcal{\ A}\otimes \mathcal{B}$ and $%
\mathrm{Tr}_{\mathcal{F}}[\upsilon ^{\dag }\upsilon ]=1$, we define a
compound state $\omega :\mathcal{A}\otimes \mathcal{B}\longrightarrow 
\mathbb{C}$ as 
\begin{equation}
\omega (A\otimes B)=\mathrm{Tr}_{\mathcal{F}}[\upsilon ^{\dag }(A\otimes
B)\upsilon ]=\mathrm{Tr}[(A\otimes B)\varpi ].
\end{equation}

\textbf{Lemma 2} ([10,11,12]) Any compound state (6) can be achieved via an
entanglement $\chi $ as 
\begin{equation}
\mathrm{Tr}_{\mathcal{G}}[\widetilde{A}\chi ^{\dag }(I\otimes B)\chi
]=\omega (A\otimes B)=\mathrm{Tr}_{\mathcal{F}\otimes \mathcal{G}}[\chi 
\widetilde{A}\chi ^{\dag }(I\otimes B)],
\end{equation}

\noindent with $\omega (A\otimes I)=\mathrm{Tr}_{\mathcal{G}}[A\varrho ]$, $%
\omega (I\otimes B)=\mathrm{Tr}_{\mathcal{H}}[B\varsigma ]$, $\widetilde{%
\varrho }=\chi ^{\dag }\chi $ and $\varsigma =\mathrm{Tr}_{\mathcal{F}}[\chi
\chi ^{\dag }]$, where $\chi $ is an operator $\mathcal{G}\longrightarrow 
\mathcal{F}\otimes \mathcal{H}$ with $\mathrm{Tr}_{\mathcal{F}}[\chi 
\mathcal{A}\chi ^{\dag }]\subset \mathcal{B}$, $\chi ^{\dag }(I\otimes
B)\chi \subset \mathcal{A}$. Moreover, the operator $\chi $ is uniquely
defined by $\widetilde{\chi }U=\upsilon $, where 
\begin{equation}
(\zeta \otimes \eta )^{\dag }\widetilde{\chi }\xi =(J\xi \otimes \eta
)^{\dag }\chi J\zeta ,\;\;\forall \xi \in \mathcal{F},\zeta \in \mathcal{G}%
,\eta \in \mathcal{H}
\end{equation}

\noindent up to a unitary transformation $U$ of the minimal space $\mathcal{F%
}=\mathrm{rank}\upsilon ^{\dag }$ equipped with an isometric involution $J$.
Note that we have used the invariance of trace under the transposition such
that $\mathrm{Tr}_{\mathcal{G}}[\widetilde{\varrho }]=\mathrm{Tr}_{\mathcal{G%
}}[\varrho ]$.

\section{Entanglement as Quantum Operation}

Quantum entanglement is iron to the classical world's bronze age. Quantum
entanglement are recently researched extensively, such as Peter Levay [15]
via geometric method, Penrose [16] and Peter Levay [17] via spinor and
twistor representation, Belavkin [10,11,12] via algebraic approach. We now
follow [10,11,12] for entangled state.

Let us write the entangled state as 
\begin{equation}
\omega (A\otimes B)=\mathrm{Tr}_{\mathcal{H}}[B\pi ^{\ast }(A)]=\mathrm{Tr}_{%
\mathcal{G}}[A\pi (B)],
\end{equation}

\noindent where the operator $\pi ^{\ast }(A)=\mathrm{Tr}_{\mathcal{F}}[\chi 
\widetilde{A}\chi ^{\dag }]\in \mathcal{B}$, bounded by $\Vert A\Vert
\varsigma \in \mathcal{B}_{\ast }$, is in the predual space $\mathcal{B}%
_{\ast }=\mathcal{T}(\mathcal{H})$ of $\mathcal{B}$ for any $A\in \mathcal{G}
$, and 
\begin{equation}
\pi (B)=J\chi ^{\dag }(I\otimes B^{\dag })\chi J=\tilde{\chi}{(I\otimes 
\widetilde{B})}\bar{\chi},
\end{equation}

\noindent with $\widetilde{B}$ defined by isometric involution in $\mathcal{H%
}$ as $\widetilde{B}=JB^{+}J$, is in $\mathcal{A}_{\ast }$ as a trace-class
operator in $\mathcal{G}$, bounded by $\Vert B\Vert \varsigma \in \mathcal{A}%
_{\ast }$.

The dual linear maps $\pi $ and $\pi ^{\ast }$ in (9), $\pi ^{\ast \ast
}=\pi $, with respect to the standard pairing $\langle A|A\rangle =\mathrm{Tr%
}[A^{\ast }A]$, are both positive, but in general not completely positive
but transpose-completely positive maps, with $\pi ^{\ast }(I)=\varsigma $, $%
\pi (I)=\varrho $.

\textbf{Remark 3} For the entangled state $\omega (A\otimes B)=\mathrm{Tr}%
[(A\otimes B)\varpi ]$, in terms of the compound density operator $\varpi
=\upsilon \upsilon ^{\dag }$, the entanglements $\pi $ and $\pi ^{\ast }$
can be written as 
\begin{equation}
\pi (B)=\mathrm{Tr}_{\mathcal{H}}[(I\otimes \widetilde{B})\varpi ],\pi
^{\ast }(A)=\mathrm{Tr}_{\mathcal{G}}[(\widetilde{A}\otimes I)\varpi ].
\end{equation}

\textbf{Definition 2} ([10,11,12]) The transpose-completely positive map $%
\pi :\mathcal{B}\rightarrow \mathcal{A}_{\ast }$ , (or its dual map $\pi
^{\ast }:\mathcal{A}\rightarrow \mathcal{B}_{\ast }$), normalized as $%
\mathrm{Tr}_{\mathcal{G}}[\pi (I)]=1$ (or, equivalently, $\mathrm{Tr}_{%
\mathcal{H}}[\pi ^{\ast }(I)]=1$) is called the quantum entanglement of the
state $\sigma (B)=\mathrm{Tr}_{\mathcal{H}}[\pi (B)]$ to a state on $%
\mathcal{A}$ described by the density operator $\varrho =\pi (I)$ (or of $%
\rho (A)=\mathrm{Tr}_{\mathcal{G}}[\pi ^{\ast }(A)]$ to $\varsigma =\pi
^{\ast }(I)$).

We call the standard entanglement $\pi =\pi _{q}$ for $(\mathcal{B},\sigma )$
the entanglement to $\varrho =\widetilde{\varsigma }$ on $\mathcal{A}=\tilde{%
\mathcal{B}}$ by 
\begin{equation}
\pi _{q}(B)=\varrho ^{1/2}\widetilde{B}\varrho ^{1/2},B\in \tilde{\mathcal{B}%
}.
\end{equation}

Obviously $\pi _{q}^{\ast }(A)=\varsigma ^{1/2}\widetilde{A}\varsigma ^{1/2}$%
, where $\varsigma =\widetilde{\varrho }$, and $\pi _{q}^{\ast }=\pi _{q}$
iff $\mathcal{B}=\tilde{\mathcal{B}}$ and $\varsigma =\widetilde{\varsigma }$%
.

The standard entanglement defines the standard compound state 
\begin{equation}
\omega _{q}(A\otimes B)=\mathrm{Tr}_{\mathcal{H}}[B\varsigma ^{1/2}%
\widetilde{A}\varsigma ^{1/2}]=\mathrm{Tr}_{\mathcal{H}}[A\varrho ^{1/2}%
\widetilde{B}\varrho ^{1/2}].
\end{equation}

\textbf{Theorem 1} Every entanglement $\pi $ on $\mathcal{B}$ to the state $%
\varrho \in \mathcal{A}_{\ast }$ has a decomposition 
\begin{equation}
\pi (B)=\sqrt{\varrho }\widetilde{\Pi (B)}\sqrt{\varrho }\equiv \pi _{q}(\Pi
(B)),
\end{equation}

\noindent where $\Pi $ is a normal completely positive map $\mathcal{B}%
\rightarrow \tilde{\mathcal{A}}$ normalized to the identity operator at
least on the minimal Hilbert subspace supporting density operator $%
\widetilde{\varrho }$. This decomposition is unique by the condition $\Pi
\left( I\right) =E_{\widetilde{\varrho }}$, where $E_{\widetilde{\varrho }%
}\in \widetilde{\mathcal{A}}$ is the orthoprojector on this minimal Hilbert
subspace $\tilde{\mathcal{G}}_{\varrho }\subseteq \mathcal{G}$.

Proof: $\Pi $ can be found as a solution to the linear equation 
\begin{equation}
\widetilde{\varrho }^{1/2}\Pi (B)\widetilde{\varrho }^{1/2}\equiv \widetilde{%
\pi (B)}\;\;\forall B\in \mathcal{B}
\end{equation}%
which is unique if $\varrho $ and therefore $\widetilde{\varrho }$ is not
degenerate: 
\begin{equation}
\Pi (B)=\widetilde{\varrho }^{-1/2}\widetilde{\pi (B)}\widetilde{\varrho }%
^{-1/2}.
\end{equation}

\noindent If $\varrho $ is degenerate, we should consider the Hilbert
subspace $\mathcal{G}_{\widetilde{\varrho }}=E_{\widetilde{\varrho }}%
\mathcal{G}$ given by the minimal orthoprojector $E_{\widetilde{\varrho }%
}\in \tilde{\mathcal{A}}$ supporting the state $\widetilde{\rho }(A)=\rho (%
\widetilde{A})$ on the transposed algebra $\tilde{\mathcal{A}}$ such that $%
\widetilde{\rho }\left( E_{\widetilde{\varrho }}\right) =1$.

\section{Quantum Mutual Information via Entanglement}

Quantum mutual information is extensively researched in the past starting
from Belavkin and Stratonovich [18] and more recently by Belavkin and Ohya
[10], Belavkin and Ohya [11], Benjamin Schumacher and Michael D.
Westmoreland [19]. Belavkin and Ohya [11,12] introduced quantum mutual
information as the von Neumann negaentropy $\mathcal{R}(\varpi )=-\mathcal{S}%
(\varpi )$ of the entangled compound state related to negaentropy $\mathcal{R%
}(\varrho \otimes \varsigma )=-\mathcal{S}(\varrho \otimes \varsigma )$ of
the product of marginal states, i.e. as the relative negaentropy $\mathcal{R}%
^{(\mathrm{a})}(\varpi :\varphi )=-\mathcal{S}^{(\mathrm{a})}(\varpi
:\varphi )$, in the sense of Lindblad, Araki and Umegaki relative entropy
[14,4,5] with respect to $\varphi =\varrho \otimes \varsigma $. Cerf and
Adami [24] discussed mutual quantum information entropy and its
subadditivity property via entropy diagram.

Note that we prefer to use in what is following the term "information" for
negaentropy, leaving the term "entropy" for the opposite quantities like
relative negainformation $\mathcal{S}^{(\mathrm{a})}(\varpi :\varphi )=-%
\mathcal{R}^{(\mathrm{a})}(\varpi :\varphi )$, which coincides with usual
von Newmann entropy $\mathcal{S}(\varpi )$ if it is taken with respect to
the trace $\phi =\mathrm{Tr}$.

We now follow [18,11] to define quantum mutual information via quantum
entanglement.

\textbf{Definition 3} Relative quantum information of Araki-Umegaki type to
compound state $\omega $ on the algebra $\mathcal{A}\otimes \mathcal{B}$,
(or information divergence of the state $\omega $ with respect to a
reference state $\phi $) is defined by the density operator $\varpi ,\varphi 
$ of these states $\omega $ and $\phi $ as 
\begin{equation}
\mathcal{R}^{(\mathrm{a})}(\varpi :\varphi )=\mathrm{Tr}[\varpi (\ln \varpi
-\ln \varphi )].
\end{equation}

This quantity is used in most definitions of quantum relative information.
However unlike the classical case, this is not only possible choice for
informational divergence of the states $\omega $ and $\phi $, and it does
not relate explicitly the informational divergence to the Radon-Nikodym type
(RN) density $\varpi _{\phi }=\varphi ^{-1/2}\varpi \varphi ^{-1/2}$ of the
state $\omega $ with respect to $\phi $ as in the classical case.

Another quantum relative information (of Belavkin-Staszewski type [26]) was
introduced in [25] as 
\begin{equation}
\mathcal{R}^{(\mathrm{b})}(\varpi :\varphi )=\mathrm{Tr}[\varpi \ln (\varphi
^{-1}\varpi )],
\end{equation}

\noindent where $\varpi \ln (\varphi ^{-1}\varpi )=\ln (\varpi \varphi
^{-1})\varpi $ is understood as the Hermitian operator 
\begin{equation}
\varpi ^{1/2}\ln (\varpi ^{1/2}\varphi ^{-1}\varpi ^{1/2})\varpi
^{1/2}=\upsilon \ln (\upsilon ^{\dagger }\varphi ^{-1}\upsilon )\upsilon
^{\dagger }.
\end{equation}

This relative information can be explicitly written in terms of the RN
density $\varpi _{\phi }$ as $\mathcal{R}^{(\mathrm{b})}(\varpi :\varphi
)=\phi (r(\varpi _{\phi }))$, where $r(\varpi _{\phi })=\varpi _{\phi }\ln
\varpi _{\phi }$.

Ohya and Petz [26] were able to show that, in finite dimensions and faithful
states, the Belavkin-Staszewski information divergence based on quantum
relative information of Belavkin and Staszewski type gives better
distinction of $\varpi $ and $\varphi $ in the sense that it is greater than
relative quantum information of Araki-Umegaki type, and that it satisfies
the following important property.

\textbf{Lemma 3} Given a normal completely positive unital map $\mathrm{K}:%
\mathcal{M}\rightarrow \mathcal{M}^{0}$, if $\omega =\omega _{0}\mathrm{K}%
,\phi =\phi _{0}\mathrm{K}$, then for both relative informations, 
\begin{equation}
\mathcal{R}(\varpi :\varphi )\leq \mathcal{R}(\varpi _{0}:\varphi _{0}).
\end{equation}

Generally this is called monotonicity property of relative information,
which is well known since [14,22] for Araki-Umegaki type, while it is less
known that Belavkin-Staszewski type also satisfies all axioms for quantum
relative entropy including this inequality. Of course it is worth
mathematically proving this inequality of Belavkin-Staszewski type in the
most general case.

\textbf{Definition 4} We define the mutual quantum information $\mathcal{I}_{%
\mathcal{A},\mathcal{B}}(\pi )=\mathcal{I}_{\mathcal{B},\mathcal{A}}(\pi
^{\ast })$ of both types in a compound state $\omega $ achieved by a quantum
entanglement $\pi :\mathcal{B}\rightarrow \mathcal{A}_{\ast }$, or by $\pi
^{\ast }:\mathcal{A}\rightarrow \mathcal{B}_{\ast }$ with 
\begin{equation}
\rho (A)=\omega (A\otimes I)=\mathrm{Tr}_{\mathcal{G}}[A\varrho ],\sigma
(B)=\omega (I\otimes B)=Tr_{\mathcal{H}}[B\varsigma ]
\end{equation}

\noindent as the relative information of each type of the state $\omega $ on 
$\mathcal{M}=\mathcal{A}\otimes \mathcal{B}$ with the respect to the product
state $\phi =\rho \otimes \sigma $: 
\begin{equation}
\mathcal{I}_{\mathcal{A},\mathcal{B}}^{(\mathrm{a})}(\pi )=\mathrm{Tr}%
[\varpi (\ln \varpi -\ln (\varrho \otimes I)-\ln (I\otimes \varsigma ))].
\end{equation}%
\begin{equation}
\mathcal{I}_{\mathcal{A},\mathcal{B}}^{(\mathrm{b})}(\pi )=\mathrm{Tr}%
[\varpi \ln ((\varrho \otimes \varsigma )^{-1}\varpi )].
\end{equation}

The definition of mutual quantum entropy for Araki-Umegaki type can be found
in [10,11,12]. Note that $\mathcal{I}_{\mathcal{A},\mathcal{B}}^{(\mathrm{a}%
)}(\pi )\leq \mathcal{I}_{\mathcal{A},\mathcal{B}}^{(\mathrm{b})}(\pi )$ as
it follows from Ohya and Petz [26].

The following inequality for Araki-Umegaki type can also be found in
[10,11,12]. Similarly this inequality for Belavkin-Staszewski type holds.

\textbf{Theorem 2} Let $\lambda :\mathcal{B}\rightarrow \mathcal{A}_{\ast
}^{0}$ be an entanglement of the state $\sigma (B)=\mathrm{Tr}[\lambda (B)]$
to $(\mathcal{A}^{0},\rho ^{0})$ with $\mathcal{A}^{0}\subseteq \mathcal{L}(%
\mathcal{G}_{0})$, $\varrho ^{0}=\lambda (I)$ on $\mathcal{B}$, and $\pi =%
\mathrm{K}_{\ast }\lambda $ be entanglement to the state $\rho =\rho ^{0}%
\mathrm{K}$ on $\mathcal{A}\subseteq \mathcal{G}$ defined as the composition
of $\lambda $ with the predual operator $\mathrm{K}_{\ast }:\mathcal{A}%
_{\ast }^{0}\rightarrow \mathcal{A}_{\ast }$ normal completely positive
unital map $\mathrm{K}:\mathcal{A}\rightarrow \mathcal{A}^{0}$. Then for
both mutual quantum informations, the following monotonicity holds 
\begin{equation}
\mathcal{I}_{\mathcal{A},\mathcal{B}}(\pi )\leq \mathcal{I}_{\mathcal{A}^{0},%
\mathcal{B}}(\lambda ).
\end{equation}

Proof: This follows from the commutativity of the following diagrams:

Applying the monotonicity property of the relative information on $\mathcal{M%
}=\mathcal{A}\otimes \mathcal{B}$ with respect to the predual map $\varpi
_{0}\mapsto (\mathrm{K}_{\ast }\otimes \mathrm{Id})(\varpi _{0})$
corresponding to $\omega _{0}\mapsto \omega _{0}(\mathrm{K}\otimes \mathrm{Id%
})$ as the ampliation $\mathrm{K}\otimes \mathrm{Id}$ of a normal completely
positive unital map $\mathrm{K}:\mathcal{A}\rightarrow \mathcal{A}^{0}$.

\textbf{Definition 5} The maximal quantum mutual information $\mathcal{J}_{%
\tilde{\mathcal{B}},\mathcal{B}}(\pi _{q})$ for both types as the supremum 
\begin{equation}
H_{\mathcal{B}}(\varsigma )=\sup_{\pi ^{\ast }(I)=\varsigma }\mathcal{I}_{%
\mathcal{B},\mathcal{A}}(\pi ^{\ast })=\mathcal{J}_{\mathcal{B},\tilde{%
\mathcal{B}}}(\pi _{q}^{\ast })
\end{equation}

\noindent over all entanglements $\pi ^{\ast }$ of any $(\mathcal{A},\rho )$
to $(\mathcal{B},\sigma )$ is achieved on $\mathcal{A}^{0}=\tilde{\mathcal{B}%
}$, $\varrho ^{0}=\tilde{\varsigma}$ by the standard quantum entanglement $%
\pi _{q}^{\ast }(A)=\varsigma ^{1/2}\widetilde{A}\varsigma ^{1/2}$ for a
fixed $\sigma (B)=\mathrm{Tr}_{\mathcal{H}}[B\varsigma ]$ is named as
entangled, or true quantum entropy of each type of the state $\sigma $.

This definition for Araki-Umegaki type can be found in [10,11,12].

\textbf{Definition 6} We call the positive difference 
\begin{equation}
H_{\mathcal{B}\mid \mathcal{A}}(\pi)=H_{\mathcal{B}}(\varsigma)-\mathcal{I}_{%
\mathcal{A},\mathcal{B}}(\pi)
\end{equation}

\noindent entangled (or true quantum) conditional entropy respectively of
each type on $\mathcal{B}$ with respect to $\mathcal{A}$.

This definition for Araki-Umegaki type can be found in [10,11,12].
Obviously, the conditional mutual quantum entropies of both types are
positive, unlike the "conditional entropies" considered for example in [8].

\section{Entangled Channel Capacity and its Additivity}

Entanglement-assisted quantum capacity, or entangled quantum capacity is
extensively researched recently, such as entangled quantum capacity
[10,11,12] and entanglement-assisted quantum capacity [20,21]. Generally C.
H. Bennett, P. W. Shor, J. A. Smolin and A. V. Thapliyal [20,21] defined
entanglement-assisted capacity of quantum channel via a common framework, we
now discuss quantum channel capacity via entanglement via mutual quantum
information entropy.

Let $\mathcal{B}\subseteq \mathcal{L}(\mathcal{H})$ be the $W^*$-algebra of
operators in a (not necessarily finite dimensional unitary) Hilbert space $%
\mathcal{H}$. Generally we denote the set of states, i.e. positive unit
trace operators in $\mathcal{B}(\mathcal{H})$ by $\mathcal{S}(\mathcal{H})$,
the set of all $m$-dimensional projections by $\mathcal{P}_m(\mathcal{H})$
and the set of all projections by $\mathcal{P}(\mathcal{H})$.

\textbf{Definition 7} A quantum channel $\Lambda $ is a normal unital
completely positive linear map (UCP) of $\mathcal{B}$ into the same or
another algebra $\mathcal{B}^{0}\subseteq \mathcal{B}(\mathcal{H}^{0})$.
These maps admit the Kraus decomposition, which is usually written in terms
of the dual map $\Lambda ^{\ast }:\mathcal{B}_{\ast }^{0}\rightarrow 
\mathcal{B}_{\ast }$ as $\Lambda ^{\ast }(\varsigma
^{0})=\sum_{k}A_{k}\varsigma ^{0}A_{k}^{\ast }\equiv \Lambda _{\ast
}(\varsigma ^{0})$ (W. F. Stinespring [13], G. Lindblad [14], A. S. Holevo,
[26]), $\Lambda (B)=\sum_{k}A_{k}^{\ast }BA_{k}$, for $A_{k}$ are operators $%
\mathcal{H}^{0}\rightarrow \mathcal{H}$ satisfying $\sum_{k}A_{k}^{\ast
}A_{k}=I^{0}$. For example, quantum noiseless channel in the case $\mathcal{B%
}=\mathcal{L}(\mathcal{H})$, $\mathcal{B}^{0}=\mathcal{L}(\mathcal{H}^{0})$
is described by a single isometric operator $Y:\mathcal{H}^{0}\rightarrow 
\mathcal{H}$ as $\Lambda (B)=Y^{\ast }BY$. See for example [22,23] for the
simple cases $\mathcal{B}=\mathcal{L}(\mathcal{H})$, $\dim (\mathcal{H}%
)<\infty $.

A noisy quantum channel sends input pure states $\sigma _{0}=\rho _{0}$ on
the algebra $\mathcal{B}^{0}=\mathcal{L}(\mathcal{H}^{0})$ into mixed states
described by the output densities $\varsigma =\Lambda ^{\ast }(\varsigma
^{0})$ on $\mathcal{B}\subseteq \mathcal{L}(\mathcal{H})$ given by the
predual $\Lambda _{\ast }=\Lambda ^{\ast }\mid \mathcal{B}_{\ast }^{0}$ to
the normal completely positive unital map $\Lambda :\mathcal{B}\rightarrow 
\mathcal{B}^{0}$ which can always be written as 
\begin{equation}
\Lambda (B)=\mathrm{Tr}_{\mathcal{F}_{+}}[Y^{\dag }BY].
\end{equation}

\noindent Here $Y$ is a linear operator from $\mathcal{H}^{0}\otimes 
\mathcal{F}_{+}$ to $\mathcal{H}$ with $\mathrm{Tr}_{\mathcal{F}%
_{+}}[Y^{\dag }Y]=I$, and ${\mathcal{F}_{+}}$ is a separable Hilbert space
of quantum noise in the channel. Each input mixed state $\sigma ^{0}$ is
transmitted into an output state $\sigma =\sigma ^{0}\Lambda $ given by the
density operator 
\begin{equation}
\Lambda ^{\ast }(\varsigma ^{0})=Y(\varsigma ^{0}\otimes I_{+})Y^{\dag }\in 
\mathcal{B}_{\ast }
\end{equation}

\noindent for each density operator $\varsigma ^{0}\in \mathcal{B}_{\ast
}^{0}$, the identity operator $I_{+}\in \mathcal{F}_{+}$.

We follow [10,11,12] to denote \QTR{cal}{\textrm{K}}$_{q}$ the set of all
normal transpose-completely positive maps $\kappa :\mathcal{A}\rightarrow 
\mathcal{B}^{0}$ with any probe algebra $\mathcal{A}$, normalized as $%
\mathrm{Tr}\kappa (I)=1$, and $\mathcal{K}_{q}(\varsigma ^{0})$ be the
subset of $\kappa \in \mathcal{K}_{q}$ with $\kappa (I)=\varsigma ^{0}$. We
take the standard entanglement $\pi _{q}^{0}$ on $(\mathcal{B}^{0},\sigma
^{0})=\left( \mathcal{A}_{0},\rho ^{0}\right) $, where $\rho _{0}(A_{0})=%
\mathrm{Tr}[A_{0}\varrho _{0}]$ given by the density operator $\varrho
_{0}=\varsigma ^{0}$, and denote by $\mathrm{K}$ a normal unital completely
positive map $\mathcal{A}\rightarrow \mathcal{A}^{0}=\widetilde{\mathcal{A}}%
_{0}$ that decomposes $\kappa $ as $\kappa (A)=\varrho _{0}^{1/2}\widetilde{%
\mathrm{K}(A)}\varrho _{0}^{1/2}$. It defines an input entanglement $\kappa
^{\ast }=\mathrm{K}_{\ast }\pi _{q}^{0}$ on the input of quantum channel as
transpose-completely positive map on $\mathcal{A}_{0}=\mathcal{B}^{0}$ into $%
\mathcal{A}_{\ast }$ normalized to $\varrho =\mathrm{K}_{\ast }\varrho ^{0}$%
, $\varrho ^{0}=\widetilde{\varrho }_{0}$.

The channel $\Lambda $ transmits this input entanglement as a true-quantum
encoding into the output entanglement $\pi =\mathrm{K}_{\ast }\pi
_{q}^{0}\Lambda \equiv \mathrm{K}_{\ast }\lambda $ mapping $\mathcal{B}$ via
the channel $\Lambda $ into $\mathcal{A}_{\ast }$ with $\pi (I)=\varrho $.
The mutual entangled information, transmitted via the channel for quantum
encoding $\kappa $ is therefore $\mathcal{J}_{\mathcal{A},\mathcal{B}%
}(\kappa ^{\ast }\Lambda )=\mathcal{J}_{\mathcal{A},\mathcal{B}}(\mathrm{K}%
_{\ast }\pi _{q}^{0}\Lambda )=\mathcal{J}_{\mathcal{A},\mathcal{B}}(\mathrm{K%
}_{\ast }\lambda )$, where $\lambda =\pi _{q}^{0}\Lambda $ is the standard
input entanglement $\pi _{q}^{0}(B)=\varsigma _{0}^{1/2}\widetilde{B}%
\varsigma _{0}^{1/2}$ with $\varsigma _{0}=\widetilde{\varsigma }^{0}$,
transmitted via the channel $\Lambda $.

\textbf{Lemma 4} Given a quantum channel $\Lambda :\mathcal{B}\rightarrow 
\mathcal{B}^{0}$, and an input state $\sigma ^{0}$ on $\mathcal{B}^{0}$, the
entangled input-output quantum information capacity via a channel $\Lambda :%
\mathcal{B}\rightarrow \mathcal{B}^{0}$ as the supremum over the set $%
\mathcal{K}_{q}(\varsigma ^{0})$ including true-quantum encodings $\kappa $
achieves the maximal value 
\begin{equation}
\mathcal{J}(\varsigma ^{0},\Lambda )=\sup_{\kappa \in \mathcal{K}%
_{q}(\varsigma ^{0})}(\kappa ^{\ast }\Lambda )=\mathcal{I}_{\mathcal{A}^{0},%
\mathcal{B}}(\lambda ),
\end{equation}

\noindent where $\lambda =\pi _{q}^{0}$ is given by the corresponding
extremal input entanglement $\pi _{q}^{0}$ mapping $\mathcal{B}^{0}=\tilde{%
\mathcal{A}^{0}}$ into $\mathcal{A}^{0}=\tilde{\mathcal{B}}^{0}$ with $%
\mathrm{Tr}[\pi _{q}(B)]=\sigma ^{0}(B)$ for all $B\in \mathcal{B}^{0}$.

Note that this Lemma for Araki-Umegaki type can be found in [10,11,12].

The following definition uses commutativity of diagrams:

\textbf{Definition 8} Given a quantum channel $\Lambda :\mathcal{B}%
\rightarrow \mathcal{B}^{0}$, and a input state $\sigma ^{0}$ on $\mathcal{B}%
^{0}$, we can define the input-output entangled information capacity as the
maximal mutual quantum information

\begin{equation}
\mathcal{J}(\varsigma ^{0},\Lambda )=\mathcal{I}_{\mathcal{B}^{0},\mathcal{B}%
}(\pi _{q}^{0}\Lambda )
\end{equation}

\noindent for input standard entanglement of the state $\varsigma ^{0}$ to
the state $\varrho ^{0}=\widetilde{\varsigma }^{0}$.

Note that this definition for Araki-Umegaki type can be found in [10,11,12].
Thus we have at least two types of such mutual quantum entropy, and
obviously, $\mathcal{J}^{(\mathrm{a})}(\varsigma ^{0},\Lambda )\leq \mathcal{%
J}^{(\mathrm{b})}(\varsigma ^{0},\Lambda )$ with input product state $\rho
_{0}^{\otimes }=\otimes _{i=1}^{n}\rho _{0}^{i}$ corresponding to the states 
$\rho _{0}^{i}=\sigma _{i}^{0}$ on $\mathcal{B}_{i}^{0}$.

Here and below for notational simplicity we implement the agreements $%
\mathcal{A}_{0}^{i}=\mathcal{B}_{i}^{0}$, $\rho _{0}^{i}=\sigma _{i}^{0}$, $%
\mathcal{A}_{0}^{\otimes }=\otimes _{i=1}^{n}\mathcal{B}_{i}^{0}$, $\rho
_{0}^{\otimes }=\otimes _{i=1}^{n}\sigma _{i}^{0}$ such that $\varsigma
_{0}^{\otimes }=\otimes _{i=1}^{n}\varrho _{i}^{0}$ is transposed input
state $\widetilde{\varrho }_{0}^{\otimes }=\otimes _{i=1}^{n}\widetilde{%
\varsigma }_{i}^{0}$ on $\mathcal{B}_{0}^{\otimes }=\otimes _{i=1}^{n}%
\mathcal{A}_{i}^{0}$ with $\widetilde{\mathcal{B}}_{i}^{0}=\mathcal{A}%
_{i}^{0}\equiv \mathcal{B}_{0}^{i}=\widetilde{\mathcal{A}}_{0}^{i}$, $%
\widetilde{\varsigma }_{i}^{0}=\varrho _{i}^{0}\equiv \varsigma _{0}^{i}=%
\widetilde{\varrho }_{0}^{i}$,.

Let $\Lambda _i$ be channels respectively from the algebra $\mathcal{B}_i$
on $\mathcal{H}_i$ to $\mathcal{B}_i^0$ on $\mathcal{H}_i^0$ for $%
i=1,2,...,n $, and let $\Lambda ^{\otimes}=\otimes_{i=1}^n \Lambda_i$ be
their tensor product.

We now show the additivity property of this entangled input-output quantum
information capcity under a given input state, using monotonicity property
(as indicated in [10,11,12] for Araki-Umegaki type).

\textbf{Theorem 3} Let $\Lambda ^{\otimes }$ be product channel from the
algebra $\mathcal{B}^{\otimes }=\otimes _{i=1}^{n}\mathcal{B}_{i}$ to $%
\mathcal{A}_{0}^{\otimes }=\otimes _{i=1}^{n}\mathcal{A}_{0}^{i}$, and let $%
\rho _{0}^{\otimes }=\otimes _{i=1}^{n}\rho _{0}^{i}$ be the tensor product
of input states $\sigma _{0}^{i}$ on $\mathcal{B}_{0}^{i}$, then

\begin{equation}
\mathcal{J}(\varrho_0 ^{\otimes},\Lambda ^{\otimes})=\sum ^n_ {i=1} \mathcal{%
J}(\varrho_0^i,\Lambda_i).
\end{equation}

\textbf{Proof}: Take $\Lambda _{i\ast }:\mathcal{B}_{i\ast }^{0}\rightarrow 
\mathcal{B}_{i\ast }$, and $\varrho _{0}^{i}\in \mathcal{B}_{i\ast }^{0}$, $%
\varsigma _{i}=\Lambda _{i\ast }(\varrho _{0}^{i})\in \mathcal{B}_{i\ast }$,
and $\mathrm{K}_{\ast }^{(n)}:\mathcal{A}_{\ast }^{\otimes }\rightarrow 
\mathcal{A}_{\ast }^{(n)}$, where $\mathcal{A}_{0\ast }^{\otimes }=\otimes
_{i=1}^{n}\mathcal{B}_{i\ast }^{0}$, but $\mathcal{A}_{\ast }^{(n)}$ is
predual to a general, not necessarily product algebra $\mathcal{A}%
^{(n)}\subseteq \mathcal{L}(\mathcal{G}^{(n)})$. For $\pi ^{(n)}=\mathrm{K}%
_{\ast }^{(n)}\pi _{q}^{0\otimes }\Lambda ^{\otimes }$, below we consider
quantum mutual information $\mathcal{I}_{\mathcal{A}^{(n)},\mathcal{B}%
^{\otimes }}(\pi ^{(n)})$ as relative quantum entropy

\begin{equation}
\mathcal{R}((\mathrm{K}_{\ast }^{(n)}\otimes \Lambda _{\ast }^{\otimes })%
\widetilde{{\varpi }}_{0}^{\otimes }:\mathrm{K}_{\ast }^{(n)}(\varsigma
_{0}^{\otimes }){\otimes }\Lambda _{\ast }^{\otimes }({\varrho }%
_{0}^{\otimes })),
\end{equation}

\noindent where $\widetilde{\varpi }_{0}^{\otimes }=\otimes _{i=1}^{n}%
\widetilde{\varpi }_{0}^{i}$ is the density operator of the standard
compound state $\otimes _{i=1}^{n}\omega _{0}^{i}$ with $\omega
_{0}^{i}(A_{i}\otimes B_{i})=\varpi _{i}^{0}(A_{i}\otimes B_{i})=\mathrm{Tr}%
[B_{i}\sqrt{\varrho _{i}^{0}}\widetilde{A}_{i}\sqrt{\varrho _{i}^{0}}]$ for $%
A_{i}\in \widetilde{\mathcal{B}}_{i}^{0},B_{i}\in \mathcal{B}_{i}^{0}$,
corresponding to $\varsigma _{i}^{0}=\varrho _{0}^{i}$.

Applying monotonicity property (Lemma 3) of quantum relative entropy to the
probe system $(\mathcal{G}^{(n)},\mathcal{A}^{(n)})$ for this given $\varrho
_{0}^{i}$ and $\Lambda _{i}$, we obtain 
\begin{equation}
\mathcal{R}((\mathrm{K}_{\ast }^{(n)}\otimes \Lambda _{\ast }^{\otimes })%
\widetilde{\varpi }_{0}^{\otimes }:\mathrm{K}_{\ast }^{(n)}(\varsigma
_{0}^{\otimes }){\otimes }\Lambda _{\ast }^{\otimes }({\varrho }%
_{0}^{\otimes }))
\end{equation}%
\begin{equation}
\leq \mathcal{R}((\mathrm{Id}^{\otimes }\otimes \Lambda ^{\otimes })%
\widetilde{\varpi }_{0}^{\otimes }:\mathrm{Id}^{\otimes }(\varsigma
_{0}^{\otimes })\otimes \Lambda _{\ast }^{\otimes }({\varrho }_{0}^{\otimes
}))
\end{equation}%
\begin{equation}
=\sum_{i=1}^{n}\mathcal{R}((\mathrm{Id}\otimes \Lambda _{i\ast })(\widetilde{%
\varpi }_{0}):\mathrm{Id}(\varsigma _{0}^{i})\otimes \Lambda _{i\ast
}(\varrho _{0}^{i}),
\end{equation}

\noindent where $\varsigma _{0}^{i}=\varrho _{i}^{0}=\widetilde{\varrho }%
_{0}^{i}$, $\varrho _{0}^{i}=\varsigma _{i}^{0}=\widetilde{\varsigma }%
_{0}^{i}$.

The suprema over $\mathrm{K}^{(n)}$ is achieved on $\mathrm{K}^{(n)}=\mathrm{%
Id}^{\otimes }$ identically mapping $\mathcal{A}^{(n)}=\otimes _{i=1}^{n}%
\mathcal{A}_{0}^{i}$ to $\mathcal{B}_{0\ast }^{\otimes }=\otimes _{i=1}^{n}%
\mathcal{B}_{0}^{i}$, where $\mathcal{B}_{0}^{i}=\widetilde{\mathcal{B}}%
_{i}^{0}$, coinciding with such $\mathcal{A}^{(n)}$ due to $\mathcal{A}%
_{0}^{i}=\widetilde{\mathcal{B}}_{i}^{0}$.

Thus $\mathcal{J}(\varrho_0 ^{\otimes},\Lambda ^{\otimes})=\sum ^n_ {i=1} 
\mathcal{J}(\varrho_0^i,\Lambda_i)$.

\textbf{Definition 9} Given a normal unital completely positive map $%
\Lambda: \mathcal{B}\rightarrow \mathcal{A}$, the suprema 
\begin{equation}
C_q(\Lambda)=\sup_{\kappa\in \mathcal{K}_q}\mathcal{I}_{\mathcal{A},\mathcal{%
B}}(\kappa^*\Lambda)=\sup_{\varsigma^0}\mathcal{J}(\varsigma^0,\Lambda)
\end{equation}

\noindent is called the quantum channel capacity via entanglement, or
q-capacity.

Note that this definition for Araki-Umegaki type can be found in [10,11,12],
there we have two types of entangled channel capacities, and obviously $%
C_{q}^{(\mathrm{a})}(\Lambda )\leq C_{q}^{(\mathrm{b})}(\Lambda )$.

\textbf{Lemma 5} Let $\Lambda (B)=Y^{\dagger }BY$ be a unital completely
positive map $\Lambda :\mathcal{B}\rightarrow \mathcal{B}^{0}$ describing a
quantum deterministic channel by an isometry $Y:\mathcal{H}^{0}\rightarrow 
\mathcal{H}$. Then 
\begin{equation}
\mathcal{J}(\varsigma ^{0},\Lambda )=H_{\mathcal{B}_{0}}(\varsigma ^{0}),
\end{equation}%
\begin{equation}
C_{q}(\Lambda )=\ln \dim \mathcal{B}^{0}.
\end{equation}

Note that this Lemma for Araki-Umegaki type can be found in [10,11,12].

Let $\Lambda ^{\otimes}$ be product channel from the algebra $\mathcal{B}
^{\otimes}=\otimes_{i=1}^n \mathcal{B}_i$ to $\mathcal{A}_0
^{\otimes}=\otimes_{i=1}^n \mathcal{B}_i^0$. The additivity problem for
quantum channel capacity via entanglement is if it is true that 
\begin{equation}
\mathcal{C}_q(\Lambda^{\otimes})=\sum ^n_ {i=1}\mathcal{C}_q(\Lambda_i).
\end{equation}

We now still follow the idea of [10,11,12] to give a proof of this
additivity property via operational approach using monotonicity property (as
indicated in [10,11,12] for Araki-Umegaki type).

\textbf{Theorem 4} Let $\Lambda ^{\otimes}$ be product channel from the
algebra $\mathcal{B} ^{\otimes}=\otimes_{i=1}^n \mathcal{B}_i$ to $\mathcal{A%
}_0 ^{\otimes}=\otimes_{i=1}^n \mathcal{B}_i^0$, then 
\begin{equation}
\mathcal{C}_q(\Lambda^{\otimes})=\sum ^n_ {i=1}\mathcal{C}_q(\Lambda_i).
\end{equation}

\textbf{Proof}: It simply follows from the additivity (31). Indeed, 
\begin{equation}
C_q(\Lambda^{\otimes})=\sup_{\kappa\in \mathcal{K}_q^{(n)}}\mathcal{I}_{%
\mathcal{A}^{(n)}, \mathcal{B}}(\kappa^*\Lambda^{\otimes})=\sup_{\varrho_0^{%
\otimes}}\mathcal{J}(\varrho_0^{\otimes},\Lambda^{\otimes})
=\sup_{\varrho_0^{\otimes}}\sum ^n_ {i=1}\mathcal{J}(\varrho_0^i,\Lambda_i)
\end{equation}

Therefore by further taking suprema over $\varrho_0^{\otimes}$ as over
independently for each $i=1,2,...,n$, thus we have 
\begin{equation}
\mathcal{C}_q(\Lambda^{\otimes})=\sum ^n_ {i=1}\sup_{\varrho_0^{\otimes}}%
\mathcal{J}(\varrho_0^i,\Lambda_i)=\sum ^n_ {i=1}\mathcal{C}_q(\Lambda_i),
\end{equation}

\noindent which is the additivity property of entangled quantum channel
capacity due to encodings via entanglement obviously.

\textbf{Remark 4} Note that there is no such additivity for the Holevo
capacity for a arbitrary channel $\Lambda :\mathcal{B}\rightarrow \mathcal{B}%
^{0}$. Indeed, this smaller, semiclassical capacity is defined as the
supremum 
\begin{equation}
C_{d}(\Lambda )=\sup_{\kappa \in \mathcal{K}_{d}}\mathcal{I}_{\mathcal{A},%
\mathcal{B}}(\kappa ^{\ast }\Lambda )
\end{equation}

\noindent over the smaller class $\mathcal{K}_{d}\subseteq \mathcal{K}_{q}$
of the diagonal [10-12] (semiclassical) encodings $\kappa :\mathcal{A}%
\rightarrow \mathcal{B}_{\ast }^{0}$ corresponding to only diagonal
(Abelian) algebras $\mathcal{A}$. This supremum cannot in general be
achieved on the standard entanglement of $\mathcal{A}^{0}=\widetilde{%
\mathcal{B}}^{0}\equiv \mathcal{B}_{0}$ if $\mathcal{A}^{0}$ is non Abelian
corresponding to the non Abelian input algebra $\mathcal{B}^{0}$. Therefore
the supremum $\mathcal{C}_{d}(\Lambda ^{\otimes })\leq \sum_{i=1}^{n}%
\mathcal{C}_{d}(\Lambda _{i})$ can be achieved not on a product Abelian
algebra $\mathcal{A}^{(n)}$ as is was in the true quantum case where we
could take $\mathcal{A}^{(n)}=\otimes _{i=1}^{n}\mathcal{B}_{0}^{i}$ with
non Abelian $\mathcal{B}_{0}^{i}=\widetilde{\mathcal{B}}_{i}^{0}$.

\section{Conclusion}

So far, continuing in this paper research on quantum channel capacity for
one-way communication via entanglement following [10,11,12,29], we treated
two types of quantum mutual information via entanglement in algebraic
approach and corresponding quantum channel capacities via entanglement in
operational approach. Using monotonicity property of quantum mutual
information of a- and b-type introduced in [10,29] we proved additivity
property of quantum channel capacities via entanglement, therefore extending
the results of V. P. Belavkin [10,10a] to products of arbitrary quantum
channel to quantum relative entropy of both Araki-Umegaki type and
Belavkin-Staszewski type.

As written in the introduction, quantum channel capacities can have several
different formulations when considering to send classical information or
quantum information, one-way or two-way communication, prior or via
entanglement, etc. in the form of different constraints on the encoding
class $\mathcal{K}$. Anyway general quantum channel capacity with different
constraints is still a big open and challenging research problem in quantum
information theory. Much more open problems can be found in [9]. There we
anticipate some research on quantum channel capacity for two-way
communication and prior or via entanglement, i.e. trading communication and
entanglement for quantum channel capacity.

Another natural problem in this direction is to compare true quantum
capacities in quantity for some interesting quantum channels with other
smaller capacities under constraints, such as Holevo capacity,
entanglement-assistant capacity, etc., and find for which channels they
coincide.

The third natural problem in this direction is to consider quantum mutual
information via entanglement and corresponding quantum channel capacities
via entanglement for $\gamma$ type since [29] studied this third and more
general quantum relative entropy in quantum information, which also meet
more natural axiomatic properties of relative entropy.

Tracing through the original research ideas on quantum mutual information
via entanglement and corresponding quantum channel capacities via
entanglement, it is easy to find that the maximal quantum information and
capacity is achieved on maximal (standard) entanglement for a given state
and on the absolutely maximal entanglement (which exists only in finite
dimensions) without constraint on the input state.

Generally how to access those capacities, using physically implementable
operations for encodings and decodings, such as in this direction of quantum
channel capacity for one-way communication via entanglement, is of course an
open problem in quantum information and quantum computation.

All those problems wait forthcoming papers in the future.

\section{References}

[1] C. H. Bennett, G. Brassard, C. Crepeau, R. Jozsa, A. Peres and W. K.
Wootters, Teleporting an unknown quantum state via dual classic and
Einstein-Podolsky-Rosen channels, Phys. Rev. Lett. 70(1993)1895-1899.

[2] A. Ekert, Quantum cryptography based on Bell's Theorem, Phys. Rev. Lett.
67(1990)661-663.

[3] R. Jozsa and B. Schumacher, A new proof of the quantum noiseless coding
theorem, J. Mod. Opt. 41(1994)2343-2350.

[4] H. Araki, Relative Entropy of states of von Neumann Algebras,
Publications RIMS, Kyoto University, 11,809(1976)

[5] H. Umegaki, Kodai Math. Sem. Rep. 14,59 (1962)

[6] L. D. Landau, and E. M. Lifschitz, Quantum Mechanics (Non-Relativistic
Theory), 3rd ed. Oxford, England: Pergamon Press, 1977.

[7] J. von Neumann, Mathematical Foundations of Quantum Mechanics,Princeton
University Press,1955.

[8] Michael A. Nielsen and Isaac L. Chuang, Quantum Computation and Quantum
Information, Cambridge University Press 2000.

[9] Peter Shor, Quantum Information Theory: Results and Open Problems, Geom.
Funct. Anal., Special Volume-GAFA2000,816-838(2000).

[10] V. P. Belavkin, On Entangled Quantum Capacity. In: Quantum
Communication, Computing, and Measurement 3. Kluwer/Plenum, 2001, 325-333.

[10a] V. P. Belavkin, On Entangled Information and Quantum Capacity, Open
Sys. and Information Dyn, 8:1-18, 2001.

[11] V. P. Belavkin, M. Ohya, Quantum Entropy and Information in Discrete
Entangled States, Infinite Dimensional Analysis, Quantum Probability and
Related Topics 4 (2001) No. 2, 137-160.

[12] V. P. Belavkin, M. Ohya, Entanglement, Quantum Entropy and Mutual
Information, Proc. R. Soc. Lond. A 458 (2002) No. 2, 209 - 231

[13] W. F. Stinespring, Proc. Amer. Math. Soc. 6, p.211(1955)

[14] G. Lindblad, Entropy, Information and Quantum Measurements, Comm. in
Math. Phys. 33, p.305-322(1973)

[15] Peter Levay, The geometry of entanglement: metrics, connections and the
geometric phase, J. Phys. A37 (2004) 1821-1842; quant-ph/0306115.

[16] R. Penrose, Report in NCG workshop in Newton Institute, Cambridge,
Sept., 2006.

[17] Peter Levay, The twistor geometry of three-qubit entanglement,
quant-ph/0403060.

[18] V. P. Belavkin, R. L. Stratonovich, Optimization of Quantum Information
Processing Maximizing Mutual Information,Radio Eng. Electron. Phys., 19 (9),
p. 1349, 1973; quant-ph/0511042.

[19] Benjamin Schumacher, and Michael D. Westmorel, Quantum mutual
information and the one-time pad, quant-ph/0604207.

[20] C. H. Bennett, P. W. Shor, J. A. Smolin and A. V. Thapliyal,
Entanglement assisted classical capacity of noisy quantum channels, Phys.
Rev. Lett., vol. 83, pp. 3081-3084, 1999.

[21] C. H. Bennett, P. W. Shor, J. A. Smolin and A. V. Thapliyal,
Entanglement-Assisted Capacity of a Quantum Channel and the Reverse Shannon
Theorem, quant-ph/0106052.

[22] A. S. Holevo, Quantum coding theorems, Russian Math. Surveys 53:6,
1295-1331, 1998; quant-ph/9808023.

[23] G. Lindblad, Quantum entropy and quantum measurements, in: Proc. Int.
Conf. on Quantum Communication and Measurement, ed. by C. Benjaballah, O.
Hirota, S. Reynaud, Lect. Notes Phys.378, 71-80, Springer-Verlag, Berlin
1991.

[24] N. Cerf and G. Adami, Von Neumann capacity of noisy quantum channels,
Phys. Rev. A 56, pp3470-3483(1997).

[25] V. P. Belavkin and P. Staszewski, C*-algebraic generalization of
relative entropy and entropy, Ann. Inst. Henri Poincare, 37, Sec. A, 51-58,
1982.

[26] M. Ohya and D. Petz, Quantum Entropy and its Use, Springer-Verlag,
Berlin, 1993.

[27] V. P. Belavkin, Quantum stochastic positive evolutions:
characterization, construction, dilation, Commun. Math. phys. 184 (1997)
533-566.

[28] A. Einstein, B. Podolsky, and N. Rosen, Can Quantum-Mechanical
Description of Physical Reality Be Considered Complete? Phys. Rev. 47,
777-780, 1935.

[29] S. J. Hammersley and V. P. Belavkin, Information Divergence for Quantum
Channels, Infinite Dimensional Analysis.In: Quantum Information and
Computing. World Scientific, Quantum Probability and White Noise
Analysis,VXIX (2006) 149-166.

[30] B. Schumacher, Sending entanglement through noisy quantum channels,
Phys. Rev. A 54, pp 2614-2628, 1996.

\end{document}